\documentstyle[aps,epsf,overcite,tighten,preprint]{revtex}

\newcommand{\dnd}[3]{\frac{\partial^{#1} #3}{\partial #2^{#1}}}

\begin{document}

\title{Dissipation of Mechanical Energy in Fused Silica Fibers}

\author{
Andri M. Gretarsson\footnote{Electronic mail: andri@phy.syr.edu} and
Gregory M. Harry\footnote{Electronic mail: gharry@phy.syr.edu}
}

\address{Department of Physics, Syracuse University, Syracuse, NY13244-1130}

\maketitle

\begin{abstract}
To determine the dissipation induced by the surface of fused silica 
fibers, we measured the quality factor of fibers having various diameters.
We measured a maximum quality factor of 21 million and
extrapolated to obtain an intrinsic quality factor for fused silica of 30 million.
Dissipation in the surface dominated at diameters less than about $1$~mm.  
We developed a method for characterizing surface-induced dissipation 
that is independent of sample geometry or mode shape.  

\end{abstract}
\newpage

\section{Introduction}
In experiments of sufficiently high sensitivity, fluctuations in the 
thermal energy of signal-band degrees of freedom
can be an important source of noise.  
These fluctuations are due to coupling between the signal-band degrees of 
freedom of the detector and the thermal bath of other degrees of freedom.
Since this coupling is also the cause of dissipation, one can 
relate the thermal noise in a detector to the dissipation of signal-band
excitations.
The relationship is quantified, for a wide class of couplings,
by the Fluctuation-Dissipation Theorem:
\begin{equation}
\label{FDT}
x^2(f)=\frac{k_BT}{\pi^2 f^2} Re\left[\frac{1}{Z(f)}\right],
\end{equation}
where $x(f)$ is the spectral density of the fluctuations,
$T$ is the temperature of the detector, and $Z^{-1}(f)$ is the 
admittance of detector excitations. \cite{Callen, Saulson:PhysRev} 
The real part of the admittance is 
proportional to the dissipation,  so to minimize the thermal noise
of a detector at a given temperature we must minimize its dissipation.

In mechanical experiments, such as gravitational wave detectors,
the relevant thermally excited degrees of freedom are actual
mechanical vibrations of detector components.  The level at which
such a detector is limited by thermal noise can therefore depend on
the internal friction of the material from which the detector components are fabricated.  
The lower the dissipation in the material, the lower the thermal noise.
Fused silica (synthetic amorphous $\mathrm{SiO_2}$) has
extremely low dissipation for audio frequency oscillations at room temperature.
Therefore, fused silica is an excellent choice of material for the fabrication
of critical components of a thermal noise limited 
detector, operating at room temperature in the audio
frequency regime.\cite{BraginskiiBook}

Many authors have reported measurements of the dissipation in fused silica. 
\cite{Fraser,Bill,Gillespie:Thesis,Lunin3,Lunin2,Litten}
However, the measured quality factors, $Q$, are usually limited by dissipation 
mechanisms other than the intrinsic dissipation of the bulk material.
\cite{PaperYinglei,Braginski,Logan}
Particularly worrisome is the effect of a structurally defective and 
chemically impure surface, such as might result from normal handling, exposure 
to the atmosphere, or differential cooling during fabrication.\cite{Lunin1}
Although the current investigations are focused on fused silica, the condition of 
the surface is also a concern for other high-$Q$ materials.
\cite{juli}

To investigate the amount of dissipation induced by
the surface, we measured the dissipation in
fused silica fibers of varying diameters, and hence varying
surface-to-volume ratios.  
In addition to providing information about the effect of the surface,
our results may be extrapolated to give an approximate value for the
dissipation in the bulk material.

\section{The Experiment}

Our fibers were hand drawn in air,
from Heraeus Co.\ ``Suprasil 2''-brand fused silica rods, using a 
natural gas flame. 
The rods had diameters between $5$ and $9~\mathrm{mm}$, and the 
fibers were drawn from the rods as supplied, with no additional surface
preparation, either before or after drawing.  

The dissipation in the fibers was measured at room temperature using the 
ringdown method. \cite{Kovalik}$^,$\cite{PaperYinglei} 
Figure~\ref{setup} shows the experimental setup.  The fibers, 
hanging  freely, were excited with a comb capacitor providing 
an oscillating electric field with a large
gradient. \cite{Cadez}  Due to the dielectric properties of fused silica, 
an oscillating force is felt by the fiber.  After a resonant
mode was excited, the field was turned off, the capacitor
grounded, and the fiber allowed to ring freely. 

The displacement of the fiber as a function of time was measured by a 
split photodiode shadow sensor. The envelope of the displacement was 
extracted and fit to the functional form 
\begin{equation}
x_0e^{-\pi f_n \phi(f_n)t}+C,
\label{dissipation angle definition}
\end{equation}
where $t$ is time, $x_0$ is the initial amplitude, $C$ is the level of sensor and 
amplifier noise, and $\phi(f_n)$ is the loss angle at the resonance frequency 
$f_n$.   (The quality factor of a resonance is equal to the reciprocal of the
loss angle; $Q=\phi(f_n)^{-1}$.)  In this experiment the dominant source 
of noise was sensor and amplifier noise.  Equation~\ref{dissipation angle definition} 
does not include a term describing seismic excitation of the fiber, which 
unlike sensor and amplifier noise, adds to the signal in 
quadrature.\cite{Rowan1}   At the 
frequencies of interest, seismic excitation of fiber resonant modes 
was not detectable above the sensor and amplifier noise, and including such 
a term in Eq.~\ref{dissipation angle definition} did not 
measurably improve the quality of the fits. 

Great care was taken to eliminate all extrinsic sources of dissipation,
or ``excess loss.''  In any measurement of fiber $Q$'s, 
there are several sources of excess loss that must
be considered:  residual gas damping, rubbing
at the clamp-fiber interface, recoil damping, and eddy-current damping.  
\cite{PaperYinglei}

The damping due to residual gas molecules in the vacuum chamber is given by
\begin{equation} 
\phi_{gas} = \frac{\bar{v}}{\pi  f d}\,
\frac{\rho_{\scriptscriptstyle gas}}{\rho_{\scriptscriptstyle fiber}},
\end{equation}
where $d$ is the fiber diameter,  $f$ is the frequency of vibration, 
$\bar{v}$ is the average speed of the gas molecules, $\rho_{\scriptscriptstyle gas}$
is the mass density of the gas, and $\rho_{\scriptscriptstyle fiber}$
is the mass density of the fiber.  Gas damping was made negligible by conducting
our measurements at pressures around $10^{-6}$~torr. 
Taking typical values for the parameters, we estimate $\phi_{gas}\approx 10^{-9}$.

To eliminate rubbing at the clamp-fiber interface we ensured that the 
fiber oscillation did not induce elastic 
deformation of the fiber material in the clamp.  This was achieved by 
drawing the fiber monolithically from a much thicker rod (i.e. leaving the fiber 
attached to the rod from which it was drawn), then clamping the rod in
a collet as shown in Fig.~\ref{setup}.\cite{Quinn} 

Recoil damping is due to coupling between the resonant modes of the fiber
and low-Q resonances of the support structure.
Resonant modes of the fiber having frequencies close 
to resonances of the supporting structure will be very strongly damped, while other 
modes will not be as strongly coupled and will have less recoil damping.
The main diagnostic for recoil damping is therefore a strong frequency dependence
of the $Q$.\cite{Logan}   
In this experiment, recoil damping was minimized by isolating the fiber resonances from the
support resonances by a structure analogous to a double pendulum.
This also serves to isolate the fiber from seismic excitation
at the frequencies of interest.
Figure~\ref{fiber designs} shows the three different double pendulum type
structures used in this experiment. The structures   
were monolithic to prevent interfacial rubbing, and in each case we measured the
dissipation in the lowest fiber only.  In the design of type~3, the lowest ``fiber''
is simply the undrawn rod, as supplied.

Eddy-current damping occurs when the oscillating fiber carries a charge, and
the motion of the charges induces eddy currents in nearby
conductors. Resistance in the conductors dissipates the
mechanical energy stored in the currents, degrading the $Q$.
The dissipation in any given fiber was not noticeably dependent on the 
arrangement of, or distance to, nearby conductors. We conclude
that eddy-current damping was negligible in our measurements. 

Finally, we note that our  fibers were drawn from fused silica
rather than fused quartz rods.
Fused quartz is fabricated from  naturally occurring SiO$_2$, while fused 
silica  is made of synthetic SiO$_2$.
Fused quartz has been measured to have intrinsic room temperature $Q$'s of 
at most a few  million. \cite{Fraser, Rowan1,Rowan2,Geppo:talk}
These values are too low for our purposes and would obscure the dissipation
induced by the surface layer. 

Figure~\ref{Q improvement} shows the variations in $Q$ measured
with and without some of the precautions mentioned above.
If the fiber is held directly in a clamp
instead of being left attached to the rod from which it was drawn,
friction due to rubbing in the clamp dominates. Without an isolation bob, recoil damping
dominates and there is strong frequency dependence.  Fibers 
clamped and hung by the rods from which they were drawn, and 
having a central isolation bob show the highest $Q$'s, with
much less frequency dependence.  In these fibers, the difference between
fused quartz and fused silica becomes apparent, with fused silica exhibiting
substantially higher $Q$.

\section{Dissipation versus Frequency for a Typical Fiber}

For each fiber, the dissipation was measured at a number of resonance
frequencies. 
For fibers of the type~1 design, the resonance frequencies agree 
well with the resonance frequencies of a beam of circular
cross section clamped at one end:
\begin{equation}
f_n=\left\{\begin{array}{ll}
\frac{\pi}{8}\sqrt{\frac{Yd^2}{\rho L^4}}(0.597)^2 & n=1 \\
\frac{\pi}{8}\sqrt{\frac{Yd^2}{\rho L^4}}(n-\frac{1}{2})^2 & n \ge 2,
\end{array} \right.
\label{resfreqs}
\end{equation}
where $Y$ is Young's modulus, $d$ is the fiber diameter, $L$ is the fiber length,
$\rho$ is the mass per unit volume, and $n=1,2\ldots$~is the mode number.
\cite{Morse}
Figure~\ref{cantilever} shows the agreement between the measured and predicted 
resonance frequencies for a typical fiber of type~1.  The mode frequencies may 
be used to calculate the diameter of the fiber, and we find good agreement
with the average diameter measured using a micrometer. 
Similarly, for the fibers of type~3, the resonance frequencies agree with those
of a free beam of circular cross section.  As expected, the resonance
frequencies of the fiber of type~2 are not well modeled by either a 
free or a clamped beam.  This is due to the large diameter of 
the excited fiber relative to the first (lowest) isolation bob which takes
some part in the motion. Because of this, the fiber of type~2 was made with an extra
isolation bob.

Figure~\ref{qvsf} shows the dissipation versus frequency for the 
fiber whose resonance frequencies are shown in Fig~\ref{cantilever}.
The graph shows a loss peak around $100$ Hz which is due to 
thermoelastic damping. \cite{Zener} Thermoelastic damping is given
by 
\begin{equation}
\phi_{therm}=\frac{Y \alpha^2 T_0}{C} \frac{\sigma f}{1+\sigma^2f^2},
\end{equation}
where $Y$ is Young's modulus, $\alpha$ is the thermal expansion coefficient, $T_0$ 
is the fiber temperature (absolute scale), $C$ is the heat capacity per unit
volume, and $f$ is the frequency of vibrations.  The constant 
$\sigma$ is  
$$\sigma=\frac{2\pi}{13.55}  \frac{C d^2}{\kappa},$$
where $d$ is the fiber diameter and $\kappa$ is the thermal conductivity.
We found good agreement between the average measured diameter and the diameter
calculated from the position of the thermoelastic damping peak.

As can be seen from the graph, the dissipation is not purely thermoelastic and
can be modeled by the form
\begin{equation}
\phi(f)=\phi_{therm}(f)+\phi_c,
\label{damping versus frequency model}
\end{equation}
where $\phi_{therm}(f)$ is thermoelastic damping and 
$\phi_c$ is a frequency-independent damping term.  The dissipation in 
most of the fibers is in good overall agreement with the form of 
Eq.~\ref{damping versus frequency model}.
(Several fibers have a small number of modes showing anomalously 
high dissipation.  This indicates an undiagnosed source of dissipation affecting 
these modes.) 

The appearance of a frequency-independent loss angle $\phi_c$ suggests the presence 
of a dissipation source whose microscopic components have a wide range of 
activation energies.\cite{duPre}
Dissipation in the  bulk material is likely 
to be of this type. 
However, a frequency-independent
term might also arise from defects or impurities in the surface layer,
or possibly from other sources. 

One way of obtaining further information as to 
the source of $\phi_c$ is to measure its dependence on the fiber diameter.
In this way it is possible to distinguish between dissipation occurring in the
bulk volume of the fiber and dissipation occurring in the surface layer.  

\section{Model of Diameter Dependence}
 
The constant loss angle $\phi_c$ is modeled as consisting of
two parts, one due to dissipation in the bulk and one due to 
dissipation in the surface layer:
\begin{equation}
\label{assumption}
\phi_c= {(\Delta E_{bulk}+\Delta E_{surf})}/{E},
\end{equation} 
where $\Delta E_{bulk}$ is the energy lost per cycle in the bulk material,  
$\Delta E_{surf}$ is the energy lost per cycle in the surface layer, and
$E$ is the total energy stored in the oscillating fiber.  

If we make the rather general assumption that $\Delta E_{surf}$ 
is proportional to the  surface area $S$ while $\Delta E_{bulk}$ 
is proportional to the volume $V$, we may write
\begin{equation}
\label{prop}
\frac{\Delta E_{surf}}{\Delta E_{bulk}} \propto \frac{S}{V}.
\end{equation}
The coefficient of proportionality depends on the ratio of
the loss angle of the surface layer
to the loss angle of the bulk material.  It also depends on the ratio of energy 
stored in the surface layer to energy stored in the bulk.  Some complication arises
because both the loss angle of the surface layer and the density of energy 
stored in the  surface layer are functions of depth.  While we can normally calculate 
the energy density from the strain profile of the mode shape, we know little
about dissipation in the surface layer.  Since we are interested in
characterizing the dissipation in the surface independently of 
the mode of oscillation or sample geometry,  we write the coefficient 
of proportionality as a product of two factors

\begin{equation}
\label{surftobulk}
\frac{\Delta E_{surf}}{\Delta E_{bulk}} =
\mu \frac{d_s}{V/S}.
\end{equation}
The geometrical factor $\mu$ depends only on the geometry of the sample and on the 
mode of oscillation, while the ``dissipation depth'' $d_s$
depends only on the strength of dissipation in the surface layer relative to the 
bulk. The  appropriate expressions for $\mu$ and $d_s$ are calculated 
in the Appendix. When the Young's modulus of the surface layer is the same as that of 
the bulk, we have
\begin{equation}
\label{dissipation depth}
d_s =\frac{1}{\phi_{bulk}}\int_0^h \phi(n)dn,
\end{equation}
where $n$ is a coordinate measuring the distance inward from the 
surface, $\phi_{bulk} \equiv \Delta E_{bulk}/E_{bulk}$ is the loss angle of the 
bulk material, $\phi(n)$ is the loss angle of the surface layer 
as a function of depth, and $h$ is the thickness of the surface.  
For our fibers, having circular cross section and oscillating in transverse
modes, we have $\mu = 2$. 

For samples with simple geometries, $\mu$ is of order
unity and the volume-to-surface ratio has the same order of magnitude as 
the minimum thickness of the sample.  When the dissipation depth 
is small compared to the minimum thickness of the sample,
the effect of the surface on the dissipation is also small. 
When the dissipation depth is greater than or on the order
of the minimum thickness of the sample,
dissipation in the surface is likely to dominate.

Since $\phi(n)$ is seldom known explicitly, a measurement of 
the dissipation depth provides a convenient way of comparing the surface 
condition  of different samples made of the same material.
Since $E \approx E_{bulk}$ we may rewrite 
Eq.~\ref{assumption} in terms of the dissipation depth, 
\begin{equation}
\label{general result}
\phi_c = \phi_{bulk}\bigl(1+ \mu \frac{d_s}{V/S} \bigr).
\end{equation}
In our case $V/S=d/4$, and the theory predicts
\begin{equation}
\label{fiber model}
\phi_c = \phi_{bulk}\bigl(1+ 8\frac{d_s}{d} \bigr).
\end{equation} 
Equation~\ref{fiber model} is the model to which we shall compare our
data on dissipation versus fiber diameter.

\section{Dissipation Versus Diameter} 

For all of the fibers, the constant loss angle $\phi_c$ was measured 
with the surface condition ``as drawn''.  
The surface of fibers drawn in a flame is  largely free from microcracks 
\cite{Doremus,Uhlman} and we tried
to avoid damaging this surface. Although care was taken 
during transport and during installation in the apparatus,  
some of the fibers did get lightly knocked against aluminum or 
glass components.  This represents the only physical contact with the fiber
surface.  

Measurements of $\phi_c$ were made in the following way. Since  $\phi(f) 
\rightarrow \phi_c$ far from the thermoelastic damping peak, a measurement of
the total loss angle $\phi$ at a frequency where thermoelastic damping is 
known to be negligible, constitutes a direct measurement of $\phi_c$. In each 
case $\phi_c$ was taken as the lowest measured value of the loss angle for a 
particular fiber. 
In most cases, measurements of $\phi$ could be taken at a sufficiently
large range of frequencies so that those modes exhibiting the lowest dissipation
gave a good approximation to the $\phi_c$ asymptote.
The three thickest fibers however, posed some problems. We were only 
able to excite two or three modes in each, and no direct 
verification could be made of the existence of a $\phi_c$ asymptote.  In 
addition, the dissipation in these fibers is very small, and
correspondingly more sensitive to excess loss.
Table~\ref{Qtab} lists the $Q$'s measured for the three thickest fibers. 
While Fiber G exhibits constant
$Q$'s, the $Q$'s of Fiber J and Fiber K are quite frequency dependent.  
This is particularly striking for the split-frequency,
third resonance mode of Fiber K, where the $Q$ changes
by a factor of $3$ within $4$ Hz.  The source of this excess loss 
is undiagnosed. 

Figure~\ref{diameter dependence} shows $\phi_c$ versus diameter for the
10 fibers measured.
Systematic errors are likely to be far larger than the uncertainty shown 
(which represents the repeatability).
The main source of systematic error is the upward bias of 
the measured dissipation due to undiagnosed sources of excess loss. 
We fit the data to Eq.~\ref{fiber model} and 
have tried to minimize the error induced by undiagnosed excess loss
by including in the fit only those fibers whose graphs of dissipation 
versus resonance frequency do not have points deviating significantly from
the form predicted by Eq.~\ref{damping versus frequency model}.  
(For example, the fiber whose dissipation versus frequency data is 
shown in Fig.~\ref{qvsf} satisfies this criterion well.) 
The fit determines $d_s$ and $\phi_{bulk}$, which have the values
\begin{eqnarray}
\label{phi_bulk value}
d_s 		&=&180 \pm 20~\mathrm{\mu m},\\
\phi_{bulk}	&=&3.3 \pm 0.3 \times 10^{-8} .
\label{d_s value}
\end{eqnarray}
The relationship between $d_s$ and other measures of surface condition
(fiber strength, surface roughness, impurities, etc.) 
is an interesting and open question.  
Further research is required to understand the dependence on 
surface preparation methods,  storage times,  manufacturing 
and handling, etc.
The above value for $\phi_{bulk}$ is consistent with the lowest 
dissipation measured in fused silica.\cite{Bill,Lunin2} 
A quality factor of approximately $3 \times 10^7$ has been 
seen in hemispherical resonators of surface-treated, 
Russian brand KS4V fused silica at $3.7$~kHz.\cite{Lunin:Private}

With knowledge of the dissipation depth, we can use Eq.~\ref{surftobulk},
to calculate the fiber diameter, $d_{eq}$, at which surface-induced dissipation 
becomes equal in importance to bulk-induced dissipation:
$$d_{eq}=8d_s \approx 1300~\mathrm{\mu m}.$$

In order to obtain an estimate for the average loss angle in the surface layer,
we model the surface layer as a homogeneous shell of
thickness $h$ having a depth-independent loss angle, \hbox{$\phi(\vec{r})\equiv\phi_{surf}$}.
 Equation~\ref{dissipation depth} gives 
\begin{equation}
d_s = h\frac{\phi_{surf}}{\phi_{bulk}}.
\end{equation}
The literature suggests several mechanisms for chemical surface damage 
penetrating to a depth of order 1 $\mathrm{\mu m}$.\cite{Doremus2} Taking $h=1~\mathrm{\mu m}$ 
and using the values given by Eqs.~\ref{phi_bulk value}~and~\ref{d_s value} we obtain
\begin{equation}
\phi_{surf} \approx 10^{-5}.
\end{equation}





\section*{Acknowledgements}

We would like to thank Peter Saulson for advice, suggestions, support, and 
careful reading of the manuscript. We also thank William Startin and Steven
Penn for useful discussions and reading the manuscript.  Thanks to Yinglei Huang
for teaching us the basics of fiber measurements and to Mark Beilby for 
helpful discussions. Additional thanks are due to Vinod Balachandran for
contributing his time in the lab during the summer of 1998, and to John Schiller who 
is extending this work by performing surface treatments on fused silica fibers.  We 
thank James Hough and the anonymous referee for pointing out the correct 
way of including seismic noise in the description of a ringdown envelope.
We would especially like to thank the glassblower to Syracuse University, 
John Chabot, who drew all the fibers used in these measurements.  
This work was supported by Syracuse University and by National Science Foundation
grant PHY-9602157.

\appendix
\section*{Form of the Geometrical Factor and the Dissipation Depth}
 
It is possible (in a continuum approximation) to define a point 
loss angle,

\begin{equation}
\label{point dissipation angle app}
\phi(\vec{r}) \equiv {\Delta \rho_{\scriptscriptstyle E}(\vec{r})}/
{\rho_{\scriptscriptstyle E}(\vec{r})},
\end{equation}
where $\vec{r}$ represents the location within the sample,
$\rho_{\scriptscriptstyle E}(\vec{r})$ is the energy density stored
at $\vec{r}$, and $\Delta \rho_{\scriptscriptstyle E}(\vec{r})$ is the
negative change in $\rho_{\scriptscriptstyle E}(\vec{r})$ per cycle.  

The bulk may then be defined as the region of points
within the sample where $\phi(\vec{r})$ is constant. 
For the purposes of our model, we assume 
that there exists a surface layer of maximum thickness $h$ where
$\phi(\vec{r})$ varies, while elsewhere in the sample $\phi(\vec{r})$ has the 
constant value $\phi_{bulk}$. 

Recalling
\begin{equation}
\rho_{\scriptscriptstyle E}(\vec{r})=\frac{1}{2}Y(\vec{r})\epsilon^2(\vec{r})
\end{equation}
where $Y(\vec{r})$ is the Young's modulus and $\epsilon(\vec{r})$ the strain amplitude,
we can write
\begin{equation}
\Delta E_{surf} = 
\frac{1}{2}\int_{\mathcal SL}  \phi(\vec{r})Y(\vec{r})\epsilon^2(\vec{r})\, d^3r 
\end{equation}
where $\mathcal SL$ is the region constituting the surface layer. 
If we make the assumption that $\phi(\vec{r})$
and $Y(\vec{r})$ may be treated as functions of depth alone,
we can write
\begin{equation}
\label{delta E in surface app}
\Delta E_{surf}=\frac{1}{2}\int_0^h \phi(n)Y(n) \, 
\left\{ \int_{\mathcal S(\mathit{n})} \epsilon^2(\vec{r}) d^2r \right\}\,   dn,
\end{equation}
where $n$ measures the distance in from the surface and ${\mathcal S(\mathit{n})}$ is
a surface of integration at depth $n$, parallel to the actual surface of the sample.
For the bulk we have
\begin{equation}
\label{delta E in bulk app}
\Delta E_{bulk} = 
\frac{1}{2}\phi_{bulk} Y_{bulk}\int_{\mathcal V}\epsilon^2(\vec{r}) d^3r
\end{equation}
where $\mathcal V$ is the region of the bulk and $Y_{bulk}$ is Young's modulus of
the bulk.

We now make the assumption that $h$ is sufficiently small that
\begin{equation}
\label{thin surf}
\epsilon(\vec{r} \in {\mathcal SL}) \approx 
\epsilon(\vec{r} \in {\mathcal S}),
\end{equation}
where ${\mathcal{S}}={\mathcal{S}}(0)$ is the actual surface of the sample. Using 
Eqs.~\ref{delta E in surface app}--\ref{thin surf} we can write
\begin{equation}
\label{surftobulk app}
\frac{\Delta E_{surf}}{\Delta E_{bulk}} =
\mu \frac{d_s}{V/S},
\end{equation}
where $V$ is the volume of the sample, $S$ is its surface area,
and $\mu$ and $d_s$ are assigned the values

\begin{equation}
\label{mu app}
\mu  = \frac{V}{S}\, \frac{\int_{\mathcal S} \epsilon^2({\vec r}) d^2r}
{\int_{\mathcal V} \epsilon^2(\vec{r}) d^3r},
\end{equation}
\begin{equation}
\label{dissipation depth app}
d_s =\frac{1}{\phi_{bulk}Y_{bulk}}\int^h_0 \phi(n)Y(n)dn.
\end{equation}
The geometrical factor $\mu$ is a dimensionless constant which depends only on 
the sample geometry and on the class of resonances excited.  The dissipation
depth $d_s$ has the dimensions of length and provides a direct measure of the total
dissipation induced by the surface layer (normalized to the dissipation in the bulk).
In uncoated samples, the dissipation
depth provides a measure of the physical and chemical damage suffered 
by the surface of the sample.  

For a transversely oscillating fiber of circular cross section, the strain amplitude
is given, in cylindrical coordinates, by
\begin{equation}
\epsilon(\vec{r}) = \dnd{2}{z}{u(z)}\,r\cos\theta,
\end{equation}
where $u(z)$ is the displacement amplitude of the fiber from equilibrium. 
Using Eq.~\ref{mu app} we immediately have
\begin{equation}
\mu =2.
\end{equation}

\pagebreak[4]

\begin{figure}
\caption{Schematic diagram of the experimental setup.  The signal from the
split photodiode shadow sensor is fed through a differential amplifier,
bandpass filter, a lock-in amplifier and computer data acquisition system.}
\label{setup}
\end{figure}

\begin{figure}
\caption{Three different designs for fiber isolation structures. 
In each case, only the dissipation in the lowest section was measured. 
In total, ten fibers were measured.  Seven were of type 1, having diameters
less than or equal to about $1050~\mathrm{\mu m}$, one was of type 2, having 
diameter of about $3500~\mathrm{\mu m}$, and two were of type 3, having 
diameters of about $4890~\mathrm{\mu m}$ and $5930~\mathrm{\mu m}$.}
\label{fiber designs}
\end{figure}

\begin{figure}
\caption{The effect of isolation techniques.
The left-hand graph shows the results for three natural fused quartz fibers
having diameters between 350 and 500 $\mathrm{\mu m}$. 
The crosses indicate a fiber of type~1 (see Fig.~\ref{fiber designs})
clamped in a collet by the rod from which it was drawn. 
The circles indicate a similar fiber clamped in a collet by the rod from which it was 
drawn,  but lacking an isolation bob. The triangles indicate a fiber detached from 
the rod from which it was drawn, lacking an isolation bob, and clamped between two 
plates.  
The right-hand graph shows the results for a single fused silica fiber of
type~1, having diameter approximately $400~\mathrm{\mu m}$.
The crosses represent the fiber clamped in a collet 
by the rod from which it was drawn.  The circles represent the same fiber, in
the same collet, but clamped and hung from the isolation bob.}
\label{Q improvement}
\end{figure}

\begin{figure}
\caption{Fit of theoretical cantilever beam resonance frequencies to the 
resonance frequencies of a typical fiber of type~1.  The data shown is
from a fiber of length $13 \pm 1$~cm and diameter $120 \pm 20~\mu$m. (The 
relatively large uncertainties are due to the taper where the fiber exits 
the central bob.)
The crosses are the measured values and the solid line is a fit of the
measured values to Eq.~\ref{resfreqs}.}
\label{cantilever}
\end{figure}

\begin{figure}
\caption{Measured $\phi$ vs. resonance frequency for a typical fiber of type~1. The circled
bars represent the measured dissipation and uncertainty. The solid line represents
the theoretical thermoelastic plus constant damping and is a fit of the measured values to
Eq.~\ref{damping versus frequency model}.  The dashed line is the thermoelastic 
term from the fit.}
\label{qvsf}
\end{figure}


\begin{figure}
\caption{Measured $\phi_c$ vs. average measured fiber diameter.  
The circled points represent  fibers whose dissipation vs. frequency graph follows
Eq.~\ref{damping versus frequency model} without anomalous points. 
The uncertainty in $\phi$ shown is the approximate repeatability (5\%). The 
uncertainty shown in the average measured diameter is an estimate for the uncertainty
induced by the fiber taper, where it exits the rod from which it was drawn.
The solid line shows a least squares fit of Eq.~\ref{fiber model} to 
the circled points. The diameter of the thinnest fiber was not measured directly
but obtained from the position of the thermoelastic damping peak.}
\label{diameter dependence}
\end{figure}

\begin{table}
\caption{Quality factors exhibited by the the three thickest fibers. 
Repeated measurements typically vary by about 5\%; this is larger
than the measurement uncertainty. The large uncertainty 
in the average measured diameter is due to the taper 
at the upper end of the fiber.}
\label{Qtab}
\end{table}

\pagebreak[4]
{\bf Table~\ref{Qtab}}\newline \newline
\vspace{4cm} \newline

\begin{table} 
\begin{tabular}{lrrrr} 
{ Sample}& Avg. diameter $\mathrm{(\mu m)}$	& Mode number 
& $f$ (Hz) &$Q$\\ 
\hline 
Fiber G$^{\mathrm{a}}$
	& $3500\pm250$	& 2 			& $732.0\pm0.1$		& $2.1 \times 10^7$\\
	&		& 3 			& $1582.5\pm0.1$	& $2.1 \times 10^7$\\
Fiber J$^{\mathrm{b}}$ 
	& $4885\pm120$	& 2$^{\mathrm{c}}$ 	& $2157.3\pm0.1$	& $1.0 \times 10^7$\\
	& 		& 2$^{\mathrm{c}}$ 	& $2167.3\pm0.1$	& $0.41 \times 10^7$\\
Fiber K$^{\mathrm{b}}$
	& $5934\pm70$	& 3$^{\mathrm{c}}$	& $1725\pm0.5$		& $0.68 \times 10^7$\\ 
	&		& 3$^{\mathrm{c}}$ 	& $1729\pm0.5$		& $2.0  \times 10^7$\\ 
	&		& 4 & $3364.0\pm0.1$	& $0.35 \times 10^7$\\ 
\end{tabular} \vspace{\baselineskip}
$^{\mathrm{a}}$Type 2.\newline 
$^{\mathrm{b}}$Type 3.\newline 
$^{\mathrm{c}}$Since fiber cross sections are not perfectly circular, mode frequencies  are split.

\end{table}

\pagebreak[4]
\begin{figure}
\begin{center}
\epsfxsize=12cm
\leavevmode
\epsfbox{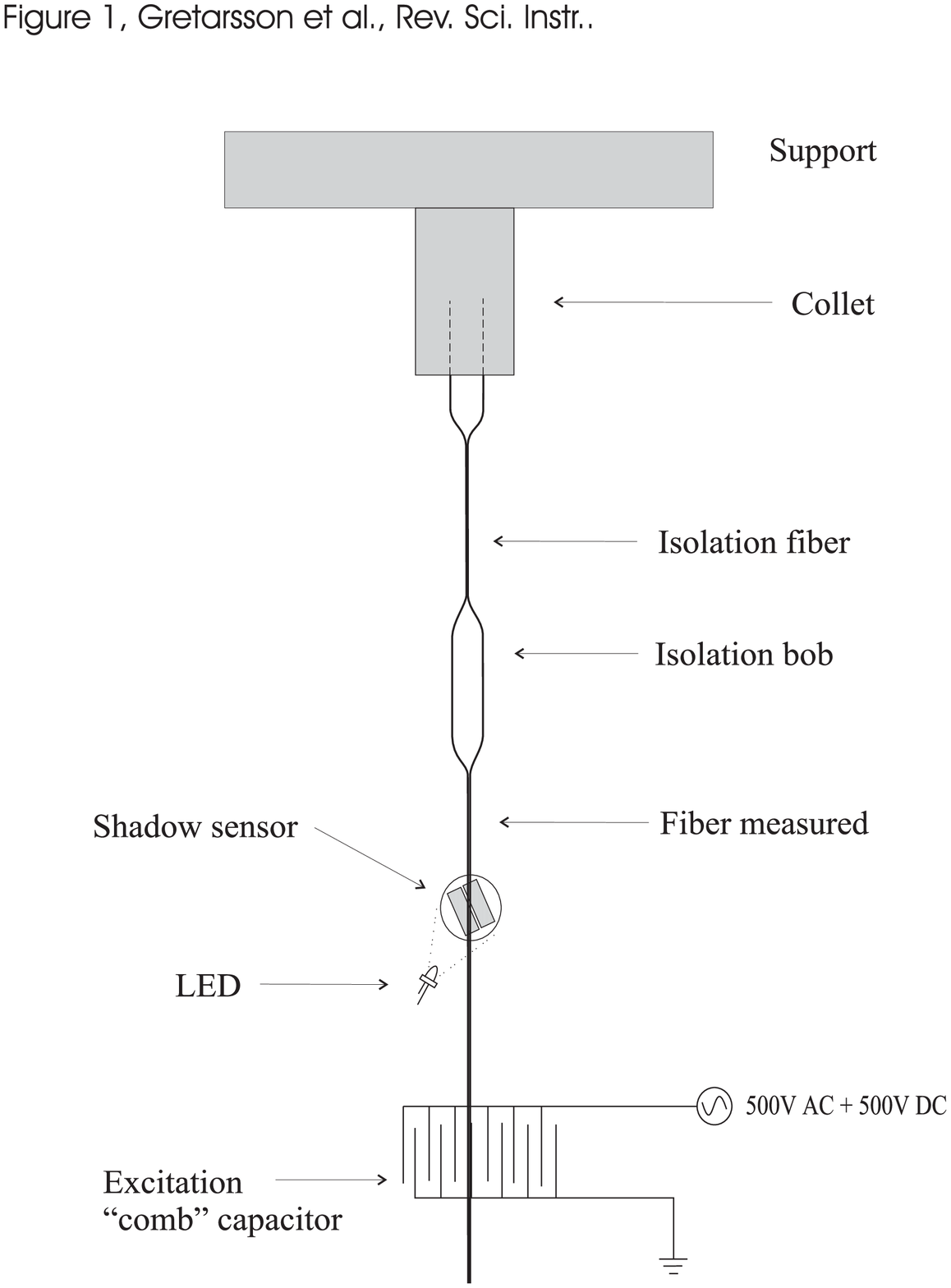}
\end{center}
\end{figure}

\pagebreak[4]
\begin{figure}
\begin{center}
\epsfxsize=12cm
\leavevmode
\epsfbox{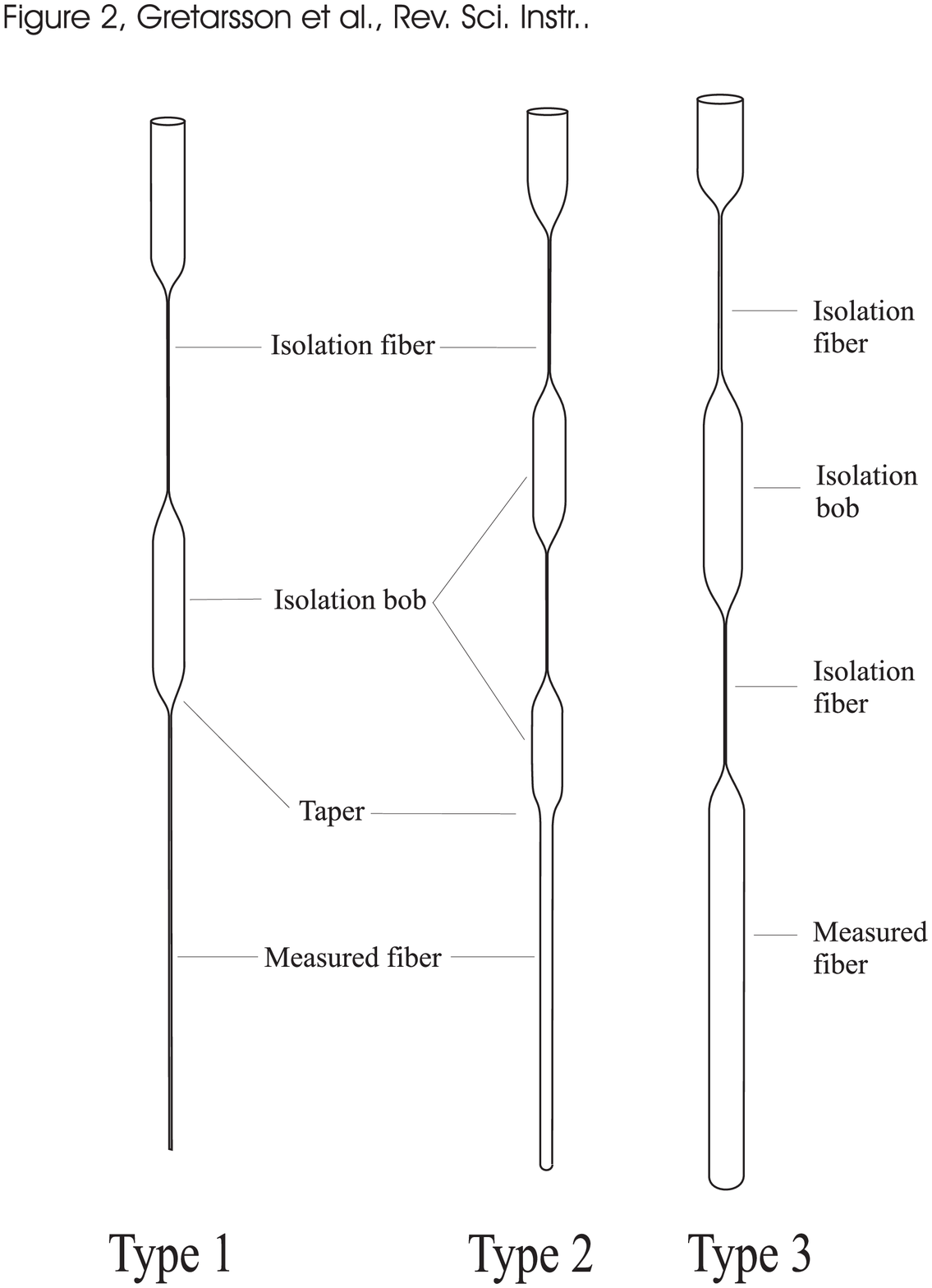} 
\end{center}
\end{figure}

\pagebreak[4]
\begin{figure}
\begin{center}
\epsfxsize=15cm
\leavevmode
\epsfbox{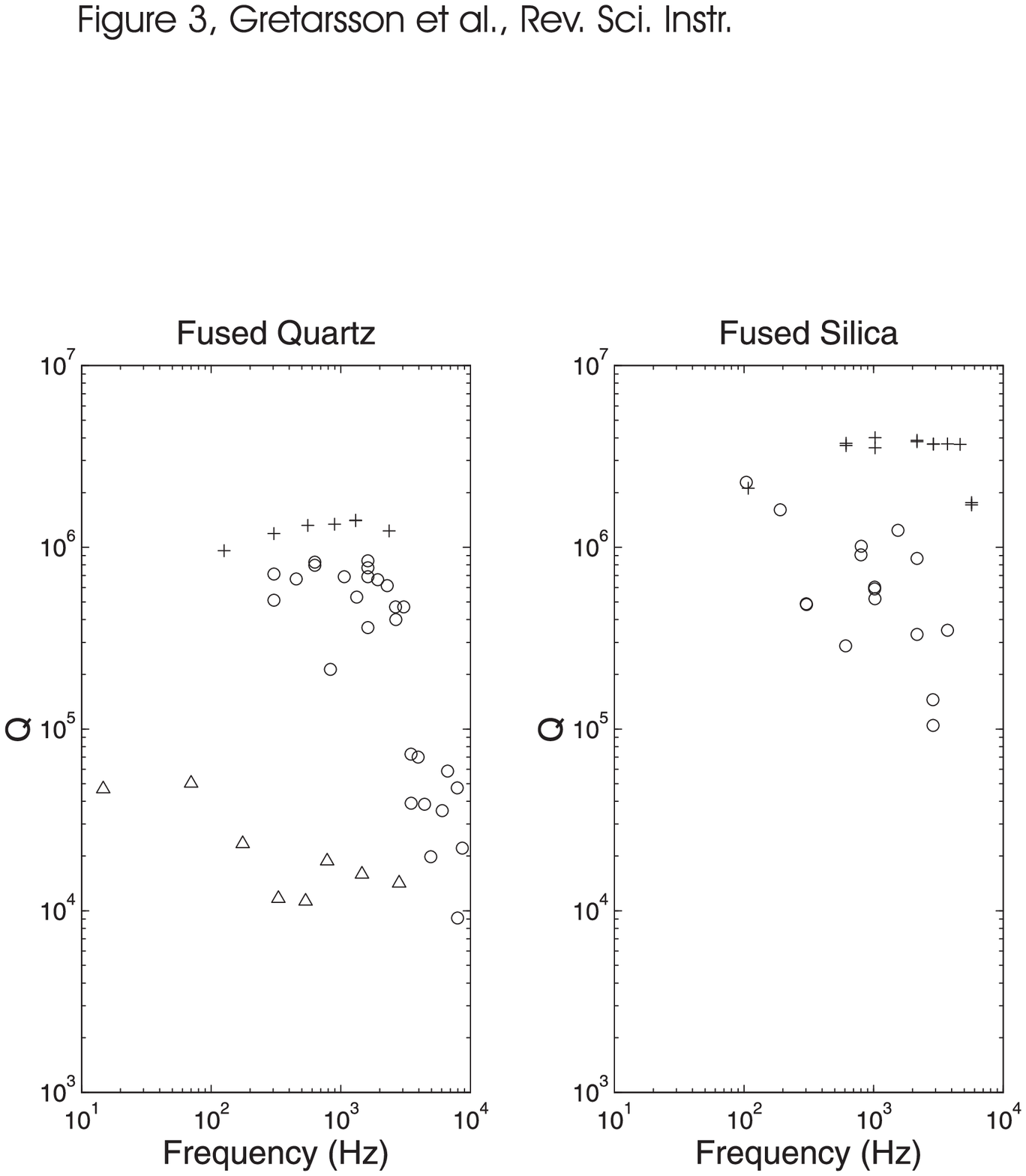}
\end{center}
\end{figure}
 
\pagebreak[4]
\begin{figure}
\begin{center}
\epsfxsize=15cm
\leavevmode
\epsfbox{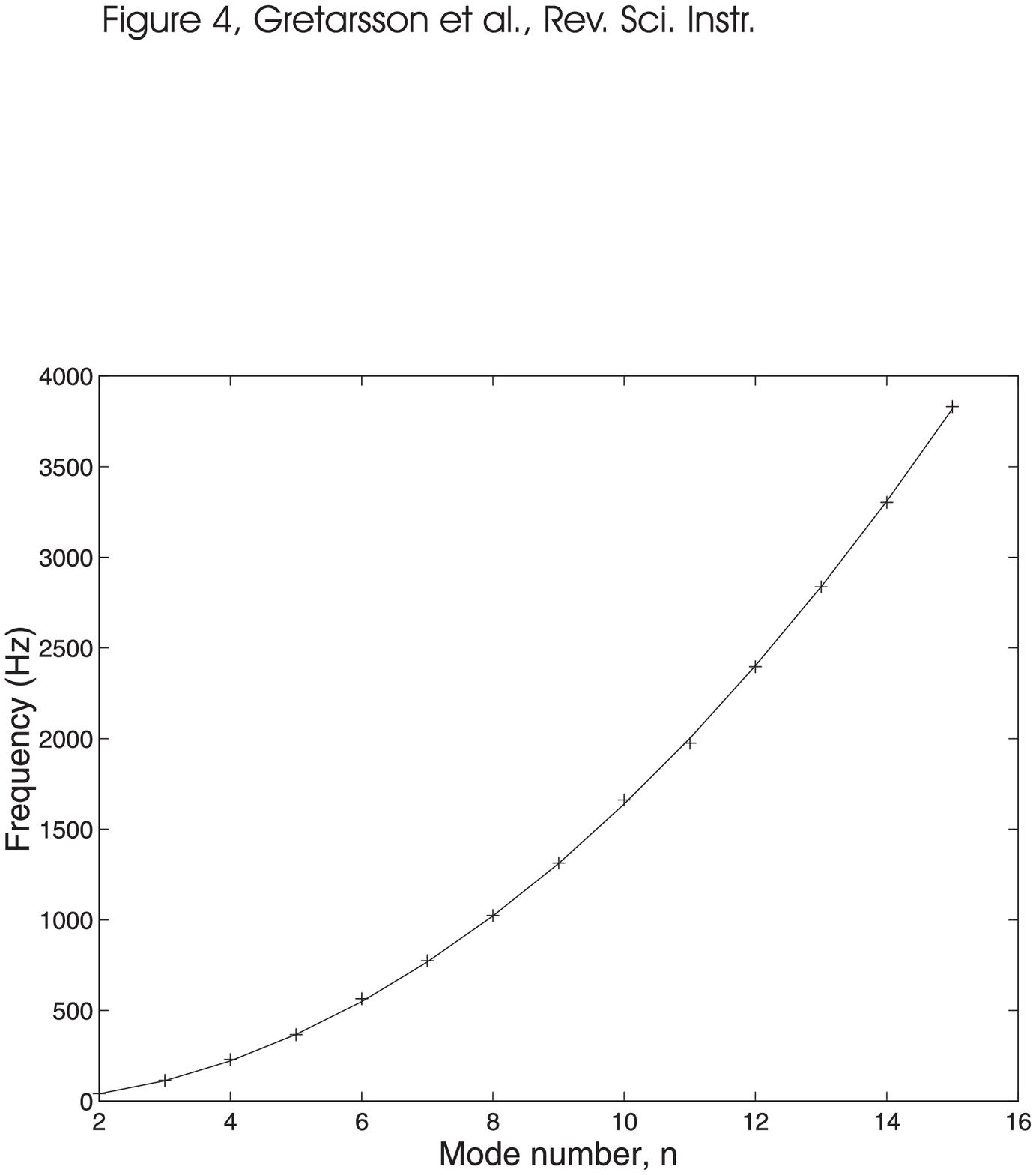}
\end{center}
\end{figure}

\pagebreak[4]
\begin{figure}
\begin{center}
\epsfxsize=15cm
\leavevmode
\epsfbox{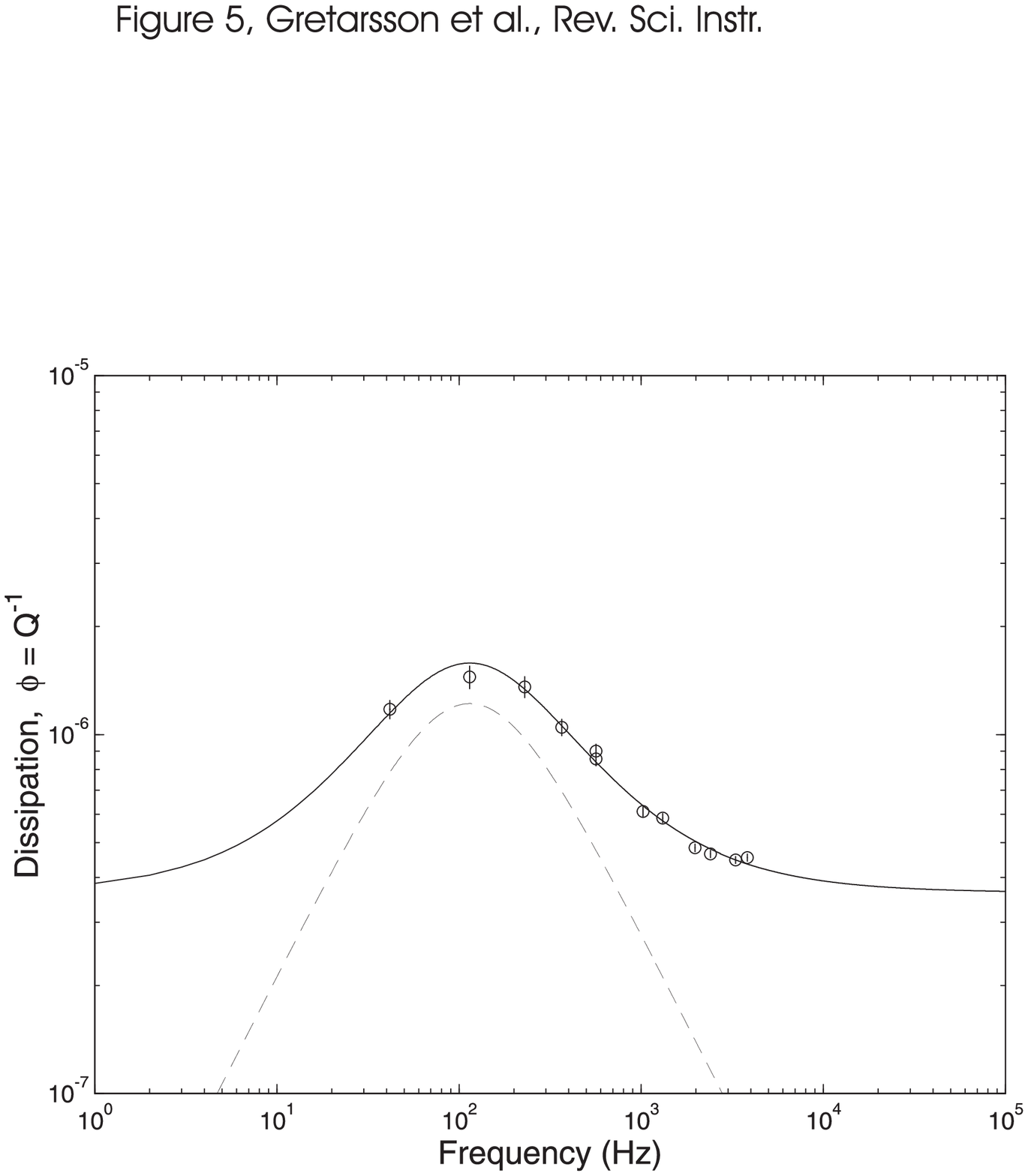}
\end{center}
\end{figure}

\pagebreak[4]
\begin{figure}
\begin{center}
\epsfxsize=15cm
\leavevmode
\epsfbox{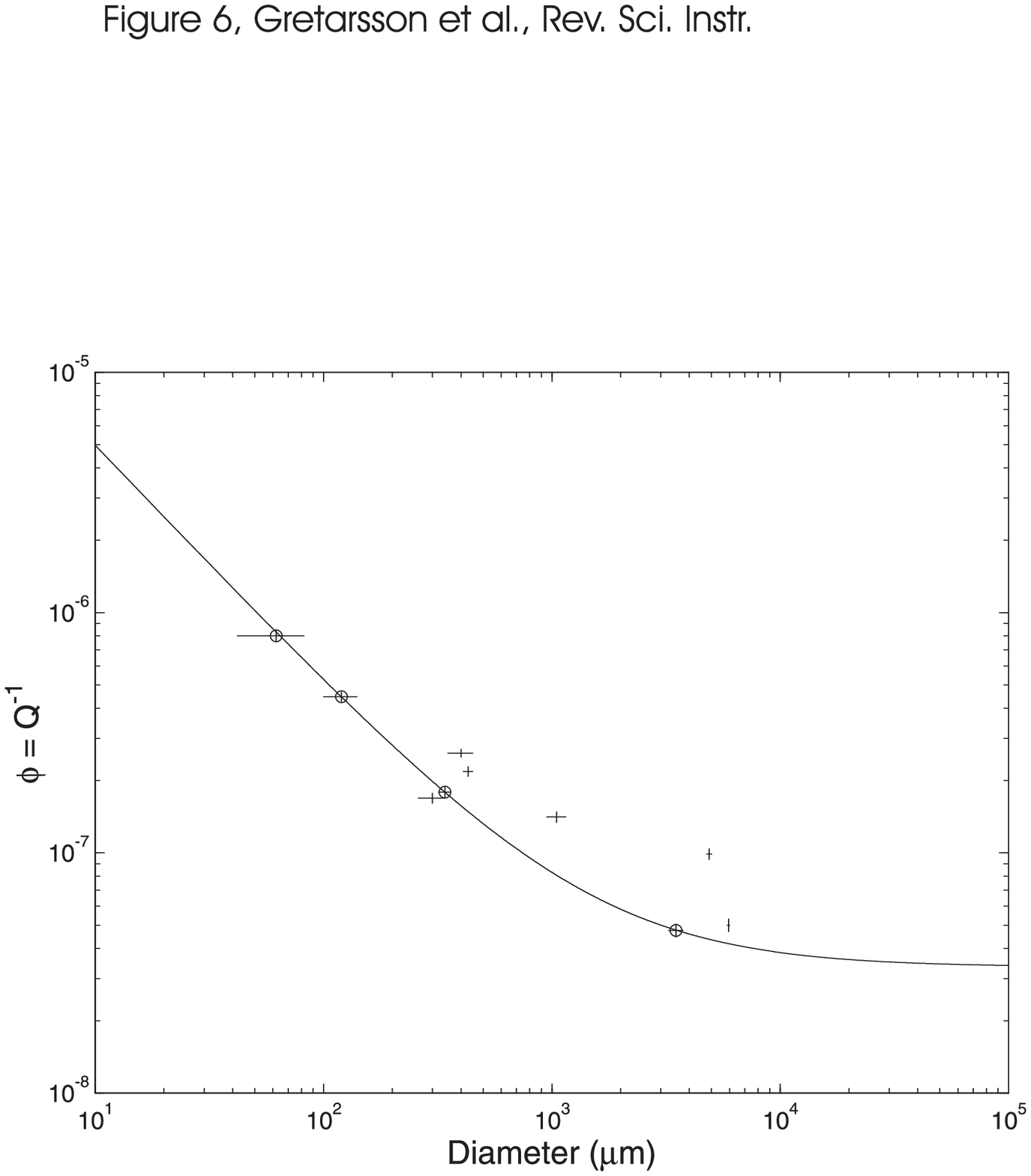}
\end{center}
\end{figure}

\end{document}